\documentclass[twocolumn,showpacs,preprintnumbers,amsmath,amssymb,prb,aps]{revtex4-1}

\usepackage{epsfig}
\usepackage{graphicx}
\usepackage{dcolumn}
\usepackage{bm}

\begin{document}

\title{Effect of pressure on the electronic and magnetic properties
of CdV$_2$O$_4$: Density functional theory studies}
\author{Sudhir K. Pandey}
\altaffiliation{Electronic mail: sk$_{_{-}}$iuc@rediffmail.com}
\affiliation{UGC-DAE Consortium for Scientific Research, University
Campus, Khandwa Road, Indore - 452001, India}
\affiliation{School of Engineering, Indian Institute of
Technology Mandi, Mandi - 175001, India}

\date{\today}

\begin{abstract}
We investigate the effect of pressure on the electronic and magnetic
states of CdV$_2$O$_4$ by using \emph{ab initio} electronic
structure calculations. The Coulomb correlation and spin-orbit
coupling play important role in deciding the structural, electronic
and magnetic properties of the compound. The total magnetic moment
of V ion is found to be $\sim$1.3 $\mu_B$ and making an angle of
$\sim$9.5 degree with the $z$-axis. In the tetragonal phase, the
ground state is the orbital ordered state where V $d_{xz}$ and
$d_{yz}$ obtitals are mainly occupied at the neighbouring sites.
This work predicts the electronic phase transition from
orbital-ordered-insulator to orbital-ordered-metal to
orbital-disordered-metal with increasing pressure. The pressure 
induced broadening of lower and upper Hubbard bands gives rise to
metal-insulator transition above 35 GPa. The simple
mean-field theory used in the present work is able to describe the
pressure dependent variation of the antiferromagnetic transition
temperature suggesting the applicability of the method in the study
of the magnetic behaviour of similar geometrically frustrated systems.

\end{abstract}


\maketitle

\section{Introduction}
The spinel vanadates having general formula AV$_2$O$_4$ (A=Cd, Mg
and Zn) offer a fertile ground to study many aspects of condensed
matter physics.
\cite{tsunetsugu,tchernyshyov,matteo,maitra,giovannetti,sudhirMVO}
At room temperature they are paramagnetic insulator with
face-centred cubic structure. The four V ions forming a regular
tetrahedron and their spins are antiferromagnetically coupled
resulting in an interesting example of a geometrically frustrated
system. At low temperature they show structural transition from
cubic to tetragonal and the tetrahedron becomes distorted as shown
in Fig. 1. The four sites of the V atoms forming the tetrahedron are
orbitally inequivalent as V1 and V2 atoms have different orbital
occupancies. The magnetic transition from paramagnetic to
antiferromagnetic (AFM) is also observed at lower
temperature.\cite{mamiya,nishiguchi,reehuis,masashige,radaelli,wheeler}

Among the above mentioned three compounds the structural transition
temperature ($T_S$) is highest ($\sim$97 K) for CdV$_2$O$_4$ and
lowest ($\sim$50 K) for ZnV$_2$O$_4$.\cite{nishiguchi,lee} On the
other hand the AFM transition temperature ($T_N$) is lowest
($\sim$35 K) for CdV$_2$O$_4$ and highest ($\sim$42 K) for
MgV$_2$O$_4$.\cite{nishiguchi,wheeler,takagi} The frustration index,
defined as $f$ $\equiv$ $|$$\theta$$_{CW}$$|$/$T_N$ where
$\theta$$_{CW}$ is the Curie-Weiss temperature and $T_N$ is the AFM
transition temperature, is considered as a measure of the activeness
of geometrical frustration.\cite{sudhirMVO,takagi,gardner} The
values of $f$ for CdV$_2$O$_4$, MgV$_2$O$_4$, and ZnV$_2$O$_4$
compounds are found to be about 11.4, 14.3, and 21.3, respectively,
suggesting that the geometrical frustration is least active in
CdV$_2$O$_4$.\cite{takagi} The pressure dependent magnetization
studies on CdV$_2$O$_4$ and ZnV$_2$O$_4$ have been carried out by
Canosa {\em et al.} and found opposite trends for the $T_N$. The
$T_N$ of CdV$_2$O$_4$ and ZnV$_2$O$_4$, respectively, increases and
decreases with increase in pressure.\cite{canosa} These contrasting
behaviours in this series of compounds are unusual and yet to be
understood.

It is clear from the above discussion that CdV$_2$O$_4$ is a
relatively simple compound where the effect of spin fluctuations due
to geometrical frustration is expected to be minimal. Thus
CdV$_2$O$_4$ would be a good starting candidate to study the complex
physical properties of the above compounds using density functional
theory (DFT), which is a mean-field theory. Here we employ the DFT
based \emph{ab inito} electronic structure calculations to study the
pressure dependent electronic and magnetic properties of
CdV$_2$O$_4$ compound. The Coulomb correlation among V 3$d$
electrons are found to be important in understanding its structural
and electronic properties. The spin-orbit coupling (SOC) is weak and
total magnetic moment of the V ion comes out to be $\sim$1.3 $\mu_B$
which is closer to the experimental value of $\sim$1.2 $\mu_B$. It
shows orbital ordered ground state where $d_{xz}$ and $d_{yz}$
orbitals are mainly occupied at the neighboring V sites. The electronic
states of the compound evolves from orbital-ordered-insulator to
orbital-ordered-metal to orbital-disordered-metal with increasing
pressure. The broadening of lower and upper Hubbard bands due to increasing pressure 
is found to be responsible for the pressure induced metal-insulator
transition. The present mean-field calculations correctly describe
the pressure dependent shift of $T_N$ indicating the applicability
of such approach in understanding the magnetic properties of similar
systems.

\section{Computational details}
The ferromagnetic (FM) and AFM calculations of CdV$_2$O$_4$ have
been carried out by using {\it state-of-the-art} full-potential
linearized augmented plane wave (FP-LAPW) method.\cite{elk} In the
AFM solution the directions of spins of V1 and V2 atoms shown in Fig.
1 are considered to be opposite. The muffin-tin sphere radii used in
the present work are 2.0, 1.85, and 1.52 Bohr for Cd, V, and O atoms,
respectively. In order to see the effect of pressure on the
electronic and magnetic properties of the compound only volume of
the primitive unit cell is varied. The atomic positions and
tetragonal distortion are kept fixed to the reported experimental
values of 85 K.\cite{onada} For the exchange correlation functional,
we used generalized gradient approximation (GGA) form of Perdew {\em
et al.}\cite{perdew} The effect of on-site Coulomb interaction among
V 3$d$ electrons is also considered within GGA+$U$ formulation of
the DFT .\cite{gga+u} The values of $U$ and $J$ used in the
calculations are 4.0 and 0.5 eV, respectively, which were found to
describe the correct electronic and magnetic properties of
MgV$_2$O$_4$ compound.\cite{sudhirMVO} The self-consistency was
achieved by demanding the convergence of the total energy to be
smaller than 10$^{-5}$ Hartree/cell. To study the role of orbital
degree of freedom in deciding the magnetic properties of the
compound the spin-orbit coupling (SOC) is also considered. The
initial axis of magnetization was set along the $z$-axis and
convergent solution was obtained in presence of SOC, which
corresponds to a non-collinear magnetism. The consideration of SOC in
the calculation is very time consuming, therefore for this solution
the energy convergence criterion was set to be smaller than
10$^{-4}$ Hartree/cell.

\section{Results and discussions}
The energy vs volume plots of CdV$_2$O$_4$ obtained from FM and AFM
GGA+$U$ calculations are show in Fig. 2. Both the curves
show almost parabolic behaviour and the volume corresponds to
minimum energy is found to be around 1092 Bohr$^3$. The energy of
AFM solution is always found to be less than that of FM solution
suggesting the AFM ground state of the compound for the range of
unit cell volumes studied here. The energy difference between FM and
AFM phase increases with decrease in volume. Thus decrease in volume
gives rise to more stability to the AFM phase in comparison to FM phase. In order to know the
exact equilibrium volume of the compound we fitted the energy-volume
data by using equation of states formula of Vinett {\em et
al.}\cite{vinett} The equilibrium volumes thus obtained correspond
to FM and AFM solutions are 1096.6 and 1093.1 Bohr$^3$,
respectively, which are about 1\% and 1.3\% less than the
experimental value of 1108.2 Bohr$^3$ at 85
K.\cite{onada,comment1} The bulk moduli of the compound found for FM
and AFM solutions are $\sim$180.8 and $\sim$170.7 GPa which is about
0.4 times less than the bulk modulus of diamond. For $U$=5 eV the
equilibrium volume and bulk modulus of the ground state AFM phase is
found to be about 1098.4 Bohr$^3$ and 175.5 GPa, respectively,
suggesting that the increased value of $U$ does not have much
influence on its structural properties. The present calculated value
of the bulk modulus is about 20 GPa more than that reported earlier
and obtained from the use of hybrid functional.\cite{canosa}

The partial density of states (PDOS) of Cd 4$d$, V 3$d$ and O 2$p$
states obtained from FM GGA+$U$ solution corresponding to
equilibrium volume are shown in Fig. 3. The valance band can be
divided in three regions: (i) region A from 0 to -2.0 eV, (ii)
region B from -2.2 to -7.8 eV and (iii) region C below -7.8 eV.
Regions A and C are dominated by V 3$d$ and Cd 4$d$ states,
respectively, whereas region B has similar contribution from all the
three atoms. The shape of Cd 4$d$, V 3$d$ and O 2$p$ PDOS are
similar in all the regions suggesting the hybridization between Cd
4$d$, V 3$d$ and O 2$p$ orbitals. The lowest lying conduction bands,
denoted by 1 and 2 in Fig. 3(b), have dominating contribution from V
3$d$ states. There is a finite gap between regions A and B which can
be attributed to the gap between bonding and antibonding bands
arising due to finite overlap of V 3$d$ and O 2$p$ orbitals. The
band gap of $\sim$0.3 eV is essentially arising due to correlation
induced splitting of V 3$d$ bands in lower and upper Hubbard bands
putting this compound into the category of Mott-Hubbard insulator.

In order to know the exact magnetic moment (MM) of the V ion we have
included SOC in the calculation and obtained the FM solution. The
spin part of MM at the V site is found to be about 1.6 $\mu_B$ and
directed along the $z$-axis. The total spin only MM is found to be 2
$\mu_B$ per formula unit (fu) suggesting that effectively V ion is
in $S$=1 spin state as expected for 3+ ionic state of the V in this
compound. The orbital moment is mainly lying in the ${yz}$-plane.
The $y$ and $z$ components of the orbital moment ($L$) are found to
be about -0.1 and -0.2 respectively where negative sign indicates
the direction of the moment. Since the direction of the spin moment
($S$) is mainly along the $z$-axis. Thus the total moment ($J$) also
lies in the ${yz}$-plane. The $y$ and $z$ components of $J$ are
found to be about -0.1 and 0.6. Using these values of $L$, $S$ and
$J$ one can estimate the magnitude and direction of total magnetic
moment of V ions. The magnitude comes out to be $\sim$1.3 $\mu_B$,
which is in good agreement with the experimental value of about 1.2
$\mu_B$.\cite{wheeler} The total magnetic moment is making an angle
of $\sim$9.5 degree with the $z$-axis. Here it is important to note
that there is a large difference between the calculated and
experimentally observed MM of MgV$_2$O$_4$ compound and this
behaviour has been attributed to the activeness of geometrical
frustration responsible for quantum
fluctuations.\cite{sudhirMVO,wheeler} However, the small difference
of 0.1 $\mu_B$ between the calculated and experimental MM for
CdV$_2$O$_4$ suggests that the geometrical frustration may be less
important for understanding its magnetic properties. This
observation can further be justified by comparing the frustration
index $f$ of both the compounds. The values of $f$ for CdV$_2$O$_4$
and MgV$_2$O$_4$ have been reported to be 11.4 and 14.3,
respectively\cite{takagi} suggesting that geometrical frustration is
less active in the CdV$_2$O$_4$. Finally one should keep in mind
that the DFT is a mean-field theory and effect of quantum
fluctuations can not be studied using this technique. Therefore, in
the geometrically frustrated system the difference between the
experimental and calculated magnetic moment is expected; and it 
can be directly related to the $f$. Since the value of $f$
for MgV$_2$O$_4$ is greater than that of CdV$_2$O$_4$, therefore the
difference between the experimental and calculated magnetic moment
for MgV$_2$O$_4$ is expected to be more in comparison to
CdV$_2$O$_4$ as mentioned above.

The electronic occupancy of $d_{xz}$ ($d_{yz}$) orbital for two
vanadium atoms V1 and V2 forming the tetrahedron (see Fig. 1) is
found to be about 0.79 and 0.06 (0.06 and 0.79), respectively. The
occupancies of rest of three V 3$d$ orbitals are same at each V
sites. This result clearly shows the orbital ordered ground state
for the CdV$_2$O$_4$ compound where $d_{xz}$ and $d_{yz}$ orbitals
are mainly occupied at V1 and V2 sites of the tetrahedron. Here it
is important to note that the AFM solution does not have much
influence on the orbital ordering and solely arising due to
tetragonal structure of the compound.

The evolution of electronic structure of the compound with
increasing pressure is shown in Fig. 4, where we have plotted total
density of states (TDOS) per formula unit (fu) obtained from AFM
solutions at different pressure. One can notice couple of obvious
changes in the TDOS with varying pressure. With increasing pressure
Cd 4$d$ states are becoming deeper in energy. The gap between
regions A and B enhances with increasing pressure suggesting the
increased energy difference between the bonding and antibonding
molecular orbitals mainly consist of V 3$d$ and O 2$p$ atomic
orbitals. This behaviour is as per the expectation because
increasing pressure increases the overlap between V 3$d$ and O 2$p$
atomic orbitals and the energy difference between bonding and
antibonding molecular orbitals is proportional to the overlap
between the atomic orbitals. With increase in pressure the band gap
decreases from $\sim$0.5 to $\sim$0.27 eV at 16.6 GPa as conduction
band (CB) 1 comes closer to the valance band (VB). The gap
almost vanishes at 34.5 GPa as the maximum of VB and minimum of CB
1 merges with each other as evident from Fig. 4(c). On further
increase of pressure the VB and CB 1 started overlapping and such
overlap of VB and CB may lead to metallic conductivity above 34.5
GPa. The TDOS/fu at the Fermi level increases from $\sim$1 states/eV
at 87.5 GPa to $\sim$2 at 105.5 GPa enhancing the conductivity by 2
times. Dispersion relations along high symmetric directions of the 
Brillouin zone for 0.23 and 34.5 GPa are shown in Fig. 5. It is evident 
from Fig. 5(a) that the valance band maximum and conduction band minimum 
lie at the $Z$ point and $\Gamma$ point, respectively. The energy difference 
between these two points gives the indirect band gap of $\sim$0.5 eV, which is responsible 
for the insulating behaviour of the compound. When pressure is increased to 34.5 GPa 
the energy difference between the $Z$ and $\Gamma$ points becomes almost zero. 
This suggests the pressure induced insulator to metal transition above 34.5 GPa.  
The increase in pressure brings V atoms closer resulting in
increased overlap between V 3$d$ orbitals. This finally leads to
broadening of upper and lower Hubbard bands giving rise to reduced
separation between them which is responsible for the decrement in the gap
with increasing pressure. Thus present work clearly indicates the
pressure induced metal-insulator transition in the compound due to
broadening of upper and lower Hubbard bands.

To study the effect of pressure on the orbital ordering we have
plotted the pressure dependent electronic occupancies of V $d_{xz}$
and $d_{yz}$ orbitals at V1 sites obtained from AFM solutions in
Fig. 6. The degree of orbital polarization at V1 sites can be
defined as $P_{orb}$ = ($d_{xz}$-$d_{yz}$)/($d_{xz}$+$d_{yz}$).
$P_{orb}$ = 1 corresponds to 100\% orbital polarization as only
$d_{xz}$ orbital is occupied at V1 sites. Needless to say that this
situation automatically tells about 100\% orbital polarization at V2
sites where only $d_{yz}$ orbital is occupied. $P_{orb}$ = 0
corresponds to orbital disordered state as same kind of orbitals
will be occupied at every sites. The occupancies of both the
orbitals show almost linear pressure dependence in the insulating
region and non-linear pressure dependence in the metallic region. 
The occupancy of $d_{xz}$
continuously decreases with increase in pressure. The pressure
dependent occupancy of $d_{yz}$ orbital shows non-monotonous
behaviour. It decreases in the insulating region and increases in
the metallic region. The $P_{orb}$ also shows non-monotonous
pressure dependence. In the insulating phase the value of $P_{orb}$
increases slowly from $\sim$0.46 at 0 GPa to $\sim$0.52 at 34.5 GPa.
However, the decrement of $P_{orb}$ is fast in the metallic phase as
its value reduces from $\sim$0.51 at 71.6 GPa to $\sim$0.32 at 105.5
GPa. This result suggests the complete melting of the orbital ordering
at higher pressure. Thus, the present work clearly shows the
interesting pressure evolution of the electronic state of
CdV$_2$O$_4$ from orbital-ordered-insulator to orbital-ordered-metal
and then to orbital-disordered-metal with increasing pressure in the
tetragonal structure.

As opposed to the electronic state, the magnetic state of the
compound is found to be robust with increasing pressure as the MM of
the V ion remains almost the same even at 105.5 GPa. In order to
know the effect of pressure on the magnetic transition temperature
we have plotted the Heisenberg exchange parameter $J$ of
neighbouring V ions as a function of pressure in Fig. 7. The
pressure dependence values of $J$ have been estimated by mapping of
energies of FM and AFM solutions to the Heisenberg Hamiltonian and
assuming that only neighbouring V ions have strong magnetic
interactions. Keeping the localized nature of V 3$d$ orbitals in
mind it may be considered as a fairly good approximation. The $J$
shows almost linear pressure dependence and it increases with
increasing pressure indicating the enhanced exchange interaction
strength due to increased overlap of V 3$d$ orbitals. If we scale
the $J$ with experimental $T_N$ then $J$ at zero pressure is
equivalent to $T_N$ at zero pressure. Using this scaling one can
plot the pressure dependence of $T_N$$(P)$/$T_N$$(0)$. For the
direct comparison with the experimental data shown in Fig. 3 of Ref.
16, we have shown the interpolated value of $T_N$$(P)$/$T_N$$(0)$ up
to 10 Kbar in the inset of Fig. 7. Our result shows about 2.5\%
increase in $T_N$ per GPa (i.e. 10 Kbar) of pressure which is closer
to the experimental observation.\cite{canosa} Keeping the complex
nature of magnetic transition in mind where spin fluctuations due to
geometrical frustrated can not be fully ignored, the applicability
of such a simple mean-field approach in understanding the pressure
dependent shift in the $T_N$ is quite interesting and demands
further investigation to such similar systems.

\section{Conclusions}
The effect of pressure on the electronic and magnetic properties of
tetragonal CdV$_2$O$_4$ have been studied by using GGA+$U$
formulation of the density functional theory. The present study
shows the importance of on-site Coulomb interaction among the V
3$d$ electrons in deriving the insulating ground state of the
compound. The spin-orbit coupling is weak and responsible for
titling the total magnetic moment ($\sim$1.3 $\mu_B$) of the V ion
by $\sim$9.5 degree with the $z$-axis. The $d_{xz}$ and $d_{yz}$
orbitals are found to be mainly occupied at neighboring V sites. The
electronic structure of the compound shows interesting evolution
from orbital-ordered-insulator to orbital-ordered-metal to
orbital-disordered-metal with increasing pressure. The pressure
induced metal-insulator transition is essentially arising due to
broadening of lower and upper Hubbard bands. The present
work clearly demonstrate the applicability of the mean-field theory
in understanding the pressure dependent behaviour of the magnetic
transition temperature in similar geometrically frustrated systems.




\section{Figure Captions:}

FIG. 1. (Color online) Atomic arrangement in the primitive unit
cell. The tetrahedron contains four V atoms denoted by V1 and V2.
V1 and V2 are orbitally inequivalent sites.

FIG. 2. (Color online) Energy vs volume plots of FM and AFM solutions
obtained from GGA+$U$ ($U$=4 eV) calculations.

FIG. 3. (Color online) Partial density of states (PDOS) for (a) Cd
4$d$, (b) V 3$d$, and (c) O 2$p$ symmetric states obtained from FM
solution. Zero in the energy axis indicates the Fermi level.

FIG. 4. (Color online) The total density of states (TDOS) per
formula unit (fu) obtained from AFM solutions at 0.23, 16.6, 34.5,
45.3, 87.5, and 105.5 GPa are shown in figures (a), (b), (c), (d),
(e), and (f), respectively. Zero in the energy axis indicates the
Fermi level.

FIG. 5. (Color online) Dispersion relations along high symmetric directions of the 
Brillouin zone for AFM solutions obtained for (a) 0.23 GPa and (b) 34.5 GPa. 

FIG. 6. (Color online) The electronic occupancies of V $d_{xz}$ and
$d_{yz}$ orbitals along with orbial polarization $P_{orb}$ =
($d_{xz}$-$d_{yz}$)/($d_{xz}$+$d_{yz}$) for V1 sites is shown as a
function of increasing pressure. The insulating and metallic regions 
are separated by vertical dashed line.

FIG. 7. (Color online) The Heisenberg exchange parameter ($J$) for
neighbouring V ions at different pressure obtained from GGA+$U$
($U$=4 eV) calculations. The inset shows the pressure dependence of
$T_N$$(P)$/$T_N$$(0)$ obtained from the scaling of the $J$. See the
text for details.




\begin{thebibliography}{99}
\bibitem{comment1} The volume quoted here is the volume of primitive unit cell containing
two formula unit whereas volume given in Ref. 18 is the volume of conventional unit
cell containing 4 formula unit.
\end{thebibliography}
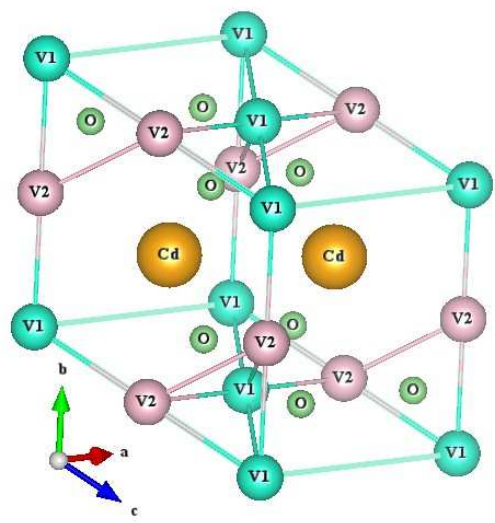
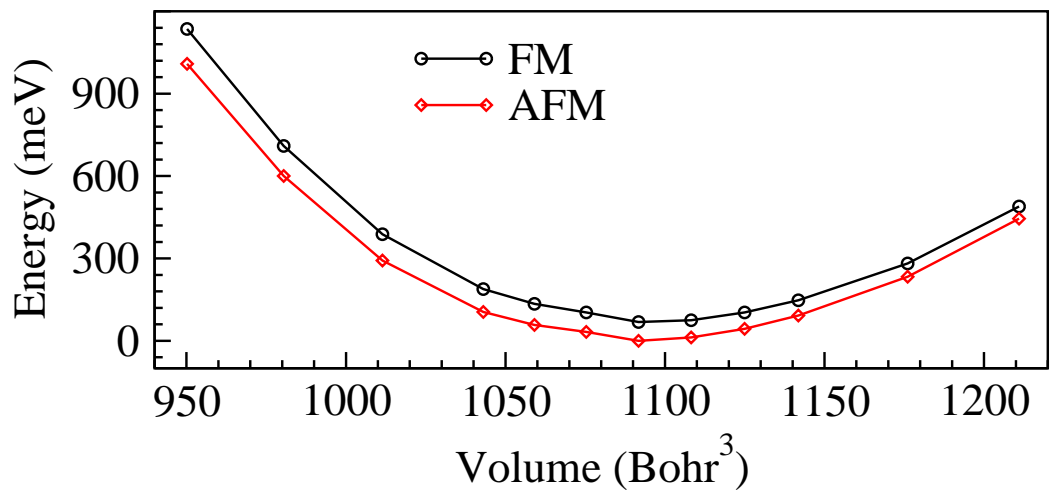
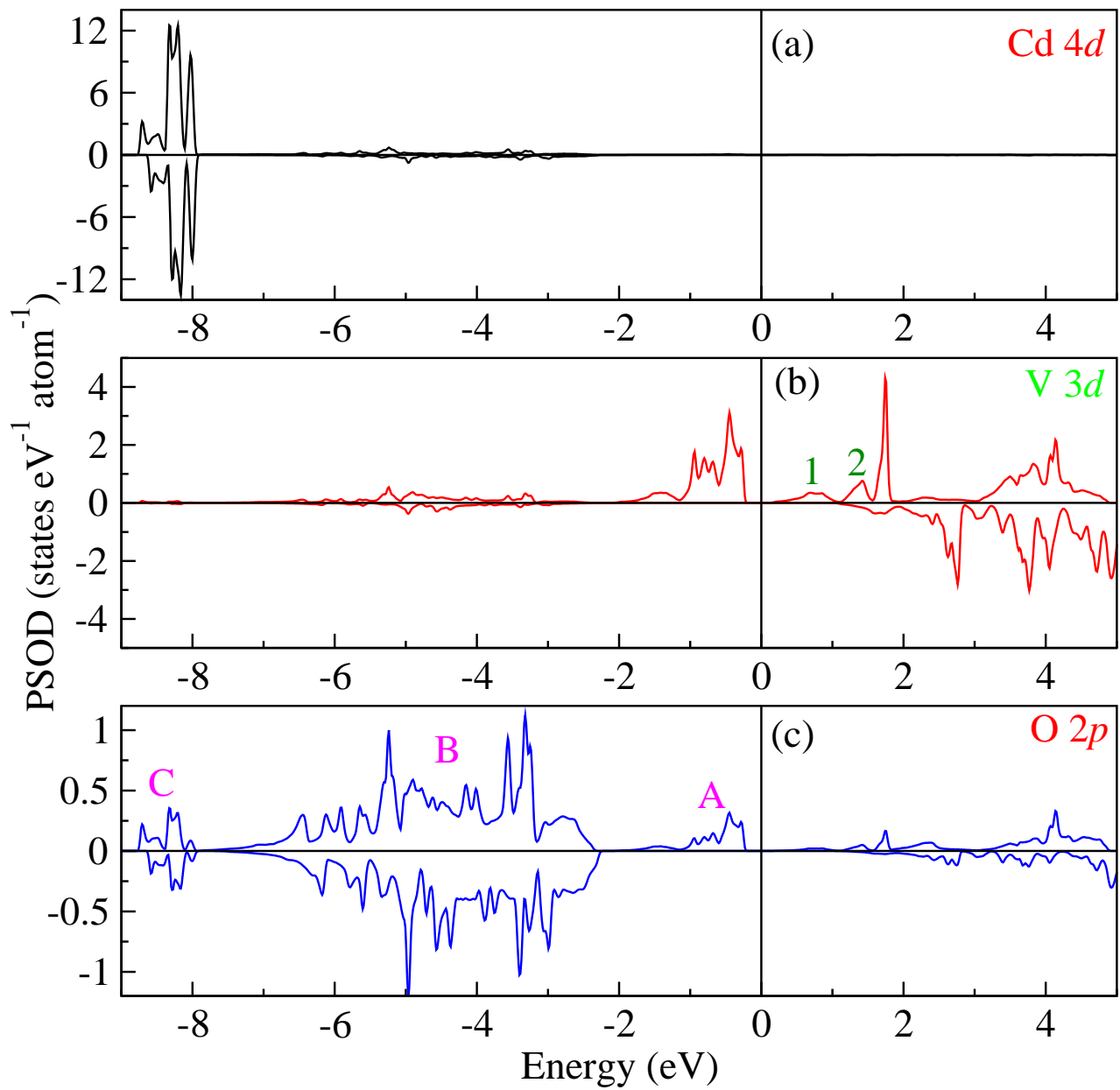
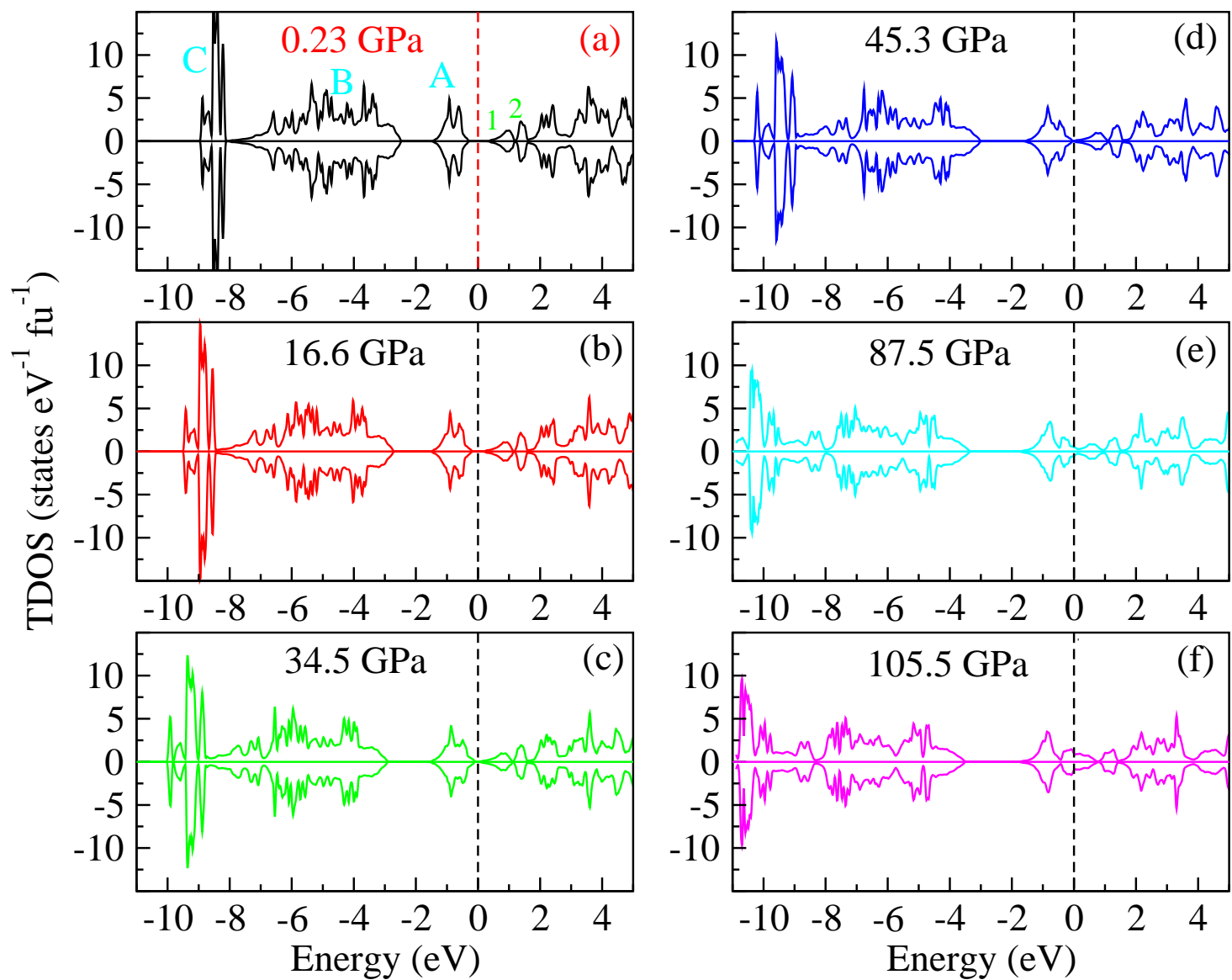
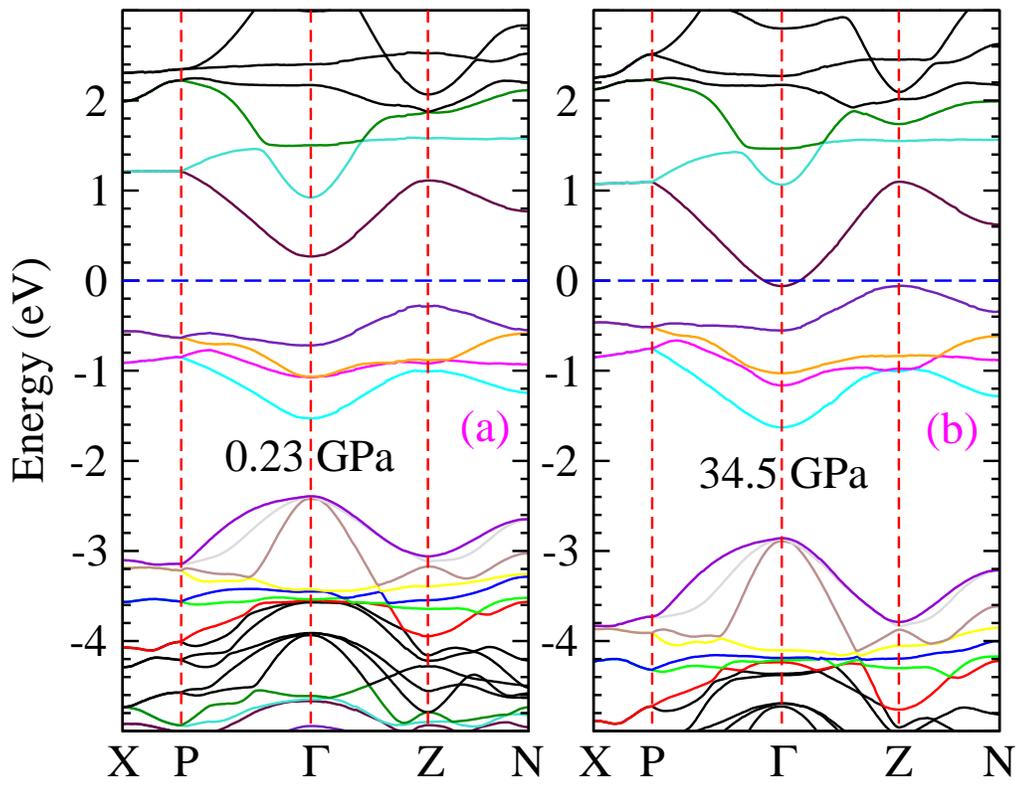
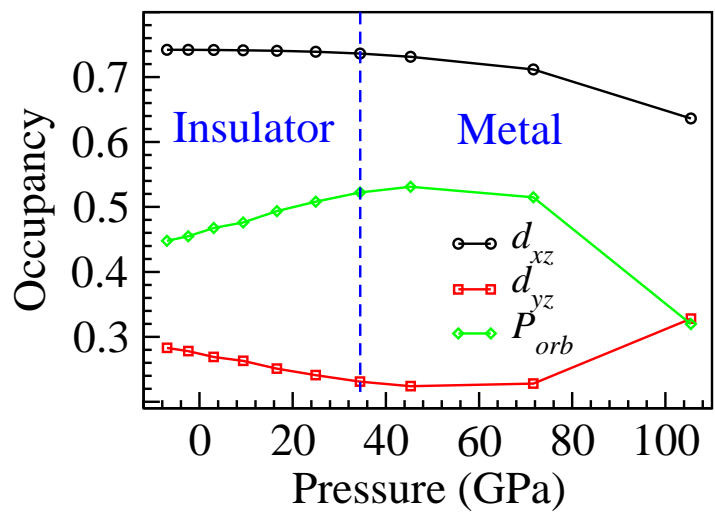
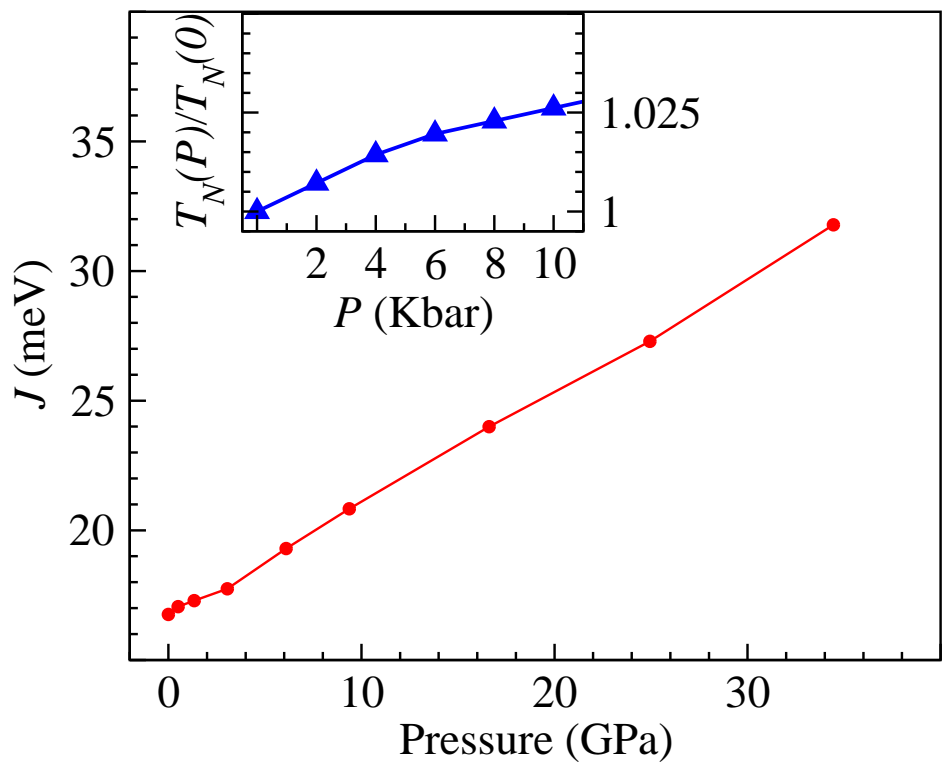
\end{document}